\documentclass[showpacs,preprintnumbers,amsmath,amssymb]{revtex4}

\usepackage[pdftex]{graphicx}
\usepackage{epstopdf}

\usepackage{dcolumn}
\usepackage{bm}

\newcommand{\be}{\begin{equation}}
\newcommand{\ee}{\end{equation}}
\newcommand{\bear}{\begin{eqnarray}}
\newcommand{\eear}{\end{eqnarray}}
\newcommand{\ba}{\begin{array}}
\newcommand{\ea}{\end{array}}

\newskip\humongous \humongous=0pt plus 1000pt minus 1000pt

\newif\ifdtup

\def\oldreffmt#1{\rlap{[#1]} \hbox to 2\parindent{}}

\def\figfmt#1{\rlap{Figure {#1}} \hbox to 1in{}}

\def\beq{\begin{equation}}
\def\eeq{\end{equation}}
\def\bea{\begin{eqnarray}}
\def\eea{\end{eqnarray}}

\def\bq{\begin{quote}}
\def\eq{\end{quote}}

\relax

\newdimen\tdim
\tdim=\unitlength
\def\bar{\overline}

\begin{document}
\setcounter{page}{0}

\begin{flushright}
ANL-HEP-PR-08-77\\ 
NUHEP-TH/08-10\\
\end{flushright}

\title{\Large  WIMPonium}

\author{\large { William Shepherd$^{a}$, Tim M.P. Tait$^{a, b, c}$, and Gabrijela Zaharijas$^{b, d}$}\\[0.5cm]
\normalsize{$^{a}$ Northwestern University, 2145 Sheridan Road, Evanston, IL 60208}\\
\normalsize{$^{b}$ Argonne National Laboratory, Argonne, IL 60439}\\
\normalsize{$^{c}$ Kavli Institute for Theoretical Physics, UCSB, CA 93106}\\
\normalsize{$^{d}$ Oskar Klein Centre for Cosmo Particle Physics, Department of Physics, Stockholm University, AlbaNova, SE-10691 Stockholm, Sweden}\\ 
}

\begin{abstract}
We explore the possibility that weakly interacting dark matter can form bound states - WIMPonium.   Such states are expected in a wide class of models of particle dark matter, including some limits of the
Minimal Supersymmetric Standard Model.  We examine the conditions under which we expect bound
states to occur, and use analogues of NRQCD applied to heavy quarkonia to
provide estimates for their properties, including couplings to the Standard Model.
We further find that it may be possible to produce WIMPonium at the LHC, and explore the properties of the
WIMP that can be inferred from measurements of the WIMPonium states.

\end{abstract}

\maketitle
\thispagestyle{empty} 

\section{Introduction}
\label{sec:intro}

The identity of the missing mass of the Universe is one of the most pressing questions at the interface of cosmology and particle physics.  The existence of dark matter is one of the few noncontroversial indications for physics beyond the Standard Model (SM), with strong preference for a weakly interacting massive particle (WIMP) as the best explanation for all of the evidence.  The hunt is on to discover WIMPs through indirect detection of its annihilation products, direct detection of its scattering with nuclei, and production at high energy accelerators.  In particular, the seeming coincidence between WIMP masses at the TeV scale and the scale of electroweak symmetry breaking would seem to argue for a connection between the two.  That connection, in turn, would suggest that WIMPs should not interact only via gravity, but could have large couplings with Standard Model fields.  The realization that the thermal prediction for the  cosmological relic density of WIMPs would roughly explain the observations lends further credence to the scenario.

While a successful WIMP should be weakly interacting with the SM, 
constraints on the interactions between the WIMPs
themselves (or among particles of the entire ``dark" sector) are quite weak.  
The relic density suggests a size for the interaction between WIMP and ordinary SM particles, but says little about the interactions among the WIMPs themselves.
Quite generally, there may be large forces between WIMPs mediated by light particles, and their exchange
can induce a Yukawa potential between WIMPs, which may be attractive.
In some cases, there will be bound states of WIMPs as a result -- ``WIMPonium".
When WIMPs are literally weakly interacting (that is, charged under the $SU(2)_L \times U(1)_Y$
of the Standard Model), they will often couple to the $Z$ boson, and perhaps also to the Higgs.  WIMPonium
is potentially a generic feature of models of dark matter.

In this article we examine the circumstances under which we expect WIMPonium to have interesting consequences, and explore some of the phenomena which are expected to result.  Early studies have focused on the specific scenario of WIMPs charged under the electroweak $SU(2) \times U(1)$ and examined the effects on the prediction for
 indirect detection signals \cite{Hisano:2003ec} or the relic density \cite{Hisano:2006nn}.
 It has also been recently suggested that WIMPonium states, either realized explicitly or manifest as nearby virtual
 states in the form of a large Sommerfeld effect,
 can reconcile observations by PAMELA and ATIC \cite{Adriani:2008zr} with a thermal relic WIMP
 \cite{ArkaniHamed:2008qn,Pospelov:2008jd,MarchRussell:2008tu} (however, see \cite{Kamionkowski:2008gj}).
We take a somewhat more generic viewpoint, and allow for general interactions between the WIMPs without recourse to a very specific model of dark matter.  We also (for simplicity) restrict ourselves to the case when the WIMP's $SU(2)$ partners (if any) are not highly degenerate in mass with the WIMP, and consider effective theories containing only the WIMP itself.

This article is organized as follows.  In Section~\ref{sec:wimponium}, we review the properties of the
bound states of the Yukawa potential, and discuss the conditions under which we expect WIMPonium
states to occur in the spectrum.  In Section~\ref{sec:couplings} we make rough estimates for WIMPonium production and decays, using the language of NRQCD \cite{Caswell:1985ui} as previously applied to
systems of heavy quarkonia \cite{Bodwin:1994jh}.  Section~\ref{sec:collider} considers the possibility that WIMPonium states could be produced at future colliders and what such states could reveal about the underlying theory of dark matter, and Section~\ref{sec:astro} explores their impact on astrophysical observations.  We conclude in Section~\ref{sec:outlook}.

\section{WIMPonium Characteristics}
\label{sec:wimponium}

The Yukawa potential between two WIMPs is characterized by its strength $\alpha_\chi$ and range $m$,
\bea
V(r) & = & - \frac{\alpha_\chi}{r} e^{-mr} ~,
\eea
where
our convention is such that $\alpha_\chi > 0$ corresponds to an attractive potential, and $\alpha_\chi <0$ 
is repulsive.  The nature (attractive or repulsive) of the potential acting on a pair of WIMPs
depends on the spin of the WIMP and that of the exchanged boson (which we will refer to as the ``mediator" particle, generically denoted by $\phi$).  As is well known, a scalar exchange leads to a universally attractive potential, whereas a vector exchange will attract particles to antiparticles, but repel pairs of particles or pairs of antiparticles.

For a weakly coupled system, the parameter $\alpha_\chi$ can be computed from the non-relativistic limit of a single particle exchange (see, {\em e.g.} \cite{Peskin:1995ev} for a pedagogical review).  For complex scalar or 
Dirac fermion WIMPs exchanging a vector, $\alpha_\chi = g^2 / 4 \pi$ where $g$ is the gauge coupling of the exchanged vector.
($g = \sqrt{g_1^2 + g_2^2}$ for a $Z$-boson exchange).   Similarly, a fermion WIMP exchanging a scalar mediator
will have $\alpha_\chi$ given by the strength of its Yukawa interaction with the scalar, $\alpha_\chi = |y|^2 / 4 \pi$.
A scalar WIMP interacting with a scalar mediator via a three-point interaction has a coupling $A$ with dimensions
of mass in natural units, and the strength of the induced Yukawa potential is given by 
$\alpha_\chi = |A|^2 / (4 \pi M^2)$ where $M$ is the mass of the WIMP.

\begin{figure}
\hspace*{-1.0cm}
\includegraphics[angle=0,scale=0.5]{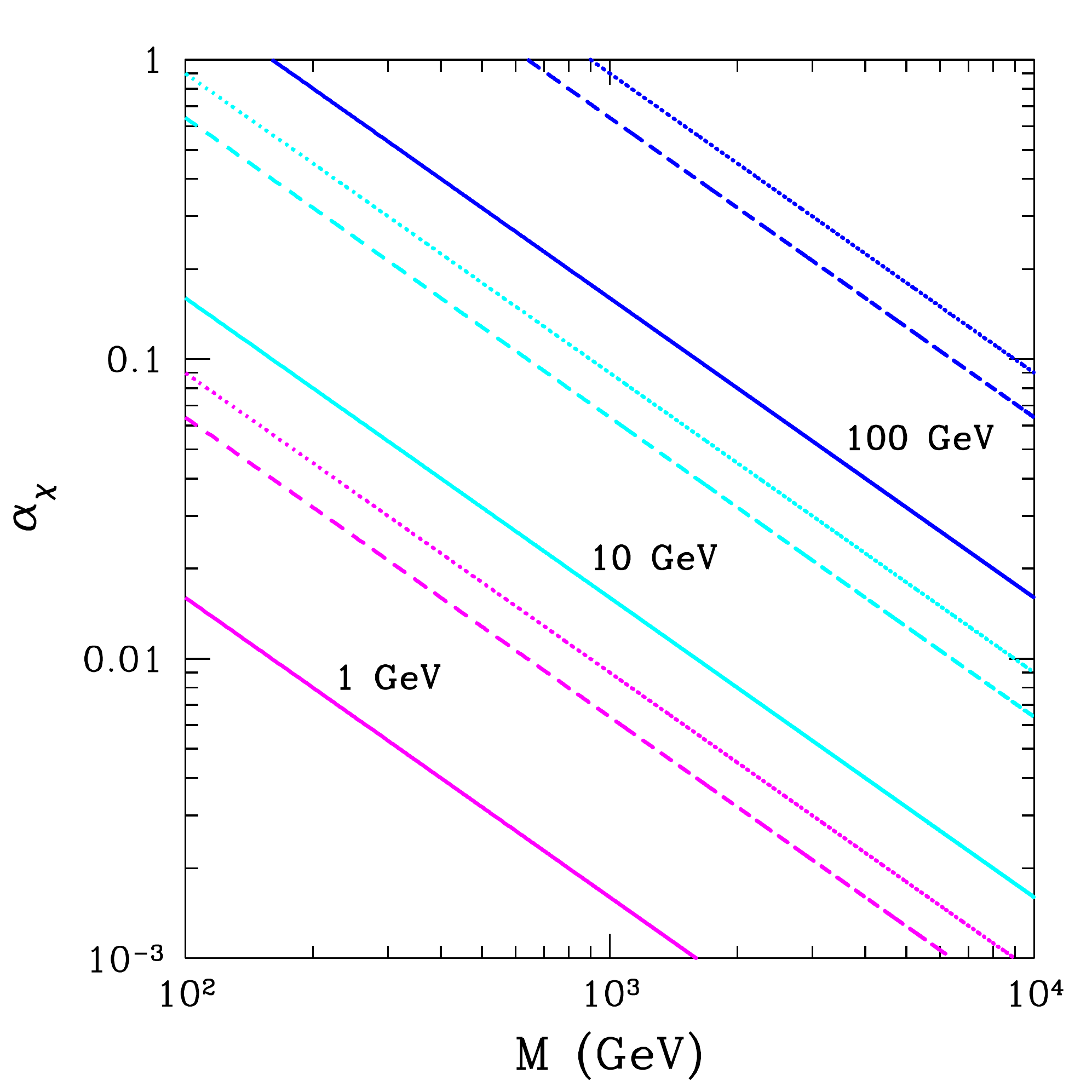} 
\caption{\label{fig:alpham}
The minimum value of $\alpha_\chi$ for which there is an $1s$  WIMPonium state (solid lines),
$2s$ states (dashed lines), and $2p$ states (dotted lines) as a function of the WIMP mass $M$
and for Yukawa screening mass (bottom curve to top curve)
$m= 1, 10$, and $100$~ GeV.
}
\end{figure}

Unlike the Coulomb potential, a non-relativistic system governed by the Yukawa potential has a finite spectrum of bound states, whose energy levels depend on the orbital angular momentum $l$.  
Detailed numerical solutions to the Schr\"odinger equation for the Yukawa potential are presented in 
\cite{Rogers:1970}, and we describe some general features of the solutions here.
The characteristic size of the system can be expressed in terms of the corresponding Bohr radius
for two WIMPs of mass $M$,
\bea
a^{-1} & = & \frac{\alpha_\chi M}{2} ~.
\eea
The system can be characterized by how ``Coulomb-like" it is through the quantity,
\bea
D & = &
\frac{1}{a m} = 
\frac{\alpha_\chi M}{2 m} ~~.
\eea
In the limit $D \gg 1$ the lowest lying states have wave functions small compared to the radius at
which the exponential suppression of the Yukawa potential is noticeable, and  their
properties are very similar to those of the Coulomb potential.  The quantity $D$ also controls the spectrum of
bound states.  In order to have a bound state at all (the $1s$ state), the potential must satisfy $D \gtrsim 0.8$.  The
$2s$ state requires $D \gtrsim 3.2$, and the $2p$ state will be present provided $D \gtrsim 4.5$.  
In figure~\ref{fig:alpham} we present the minimum $\alpha_{\chi}$ for which $1s$, $2s$, and $2p$ bound states exist as a function of the WIMP mass $M$ for several values of mediator mass $m$.

The mass of the WIMPonium state is $M_{\Psi} = 2M - E_B$ where $E_B$ is the binding energy,
\bea
E_B (n, l) &=& \frac{\alpha_\chi^2 M}{4} \epsilon (n,l) ~.
\label{eq:EB}
\eea
The function $\epsilon(n,l)$ asymptotes to $1/n^2$ in the limit $D \gg 1$.  For the $1s$ state, 
$\epsilon= \{ 0.02, 0.1, 0.8 \}$ for $D= \{1, 1.4, 10 \}$.  As expected in a weakly bound system,
the binding energy is typically quite small compared to the mass of the WIMPs themselves;
for a WIMP of mass 1 TeV with $\alpha_\chi \sim 10^{-2}$ and $D \gg 1$, the binding energy of the $1s$ state
will be $E_B \sim 25$~MeV.  

While limits on $\alpha_\chi$ are very weak or non-existent, we do
expect WIMPs to be reasonably weakly coupled, and we can generically expect that there will be a limited number of WIMPonium states.  The lowest states are the $1s$, the $2s$, and the $2p$ states.  Since the $1s$ and $2s$ states
possesses no orbital angular momentum, the spin of the WIMPonium will be determined entirely by the spins of the WIMPs themselves.  As symmetric states, they further require that the spins of the WIMPs enforce either
Bose-Einstein or Fermi-Dirac statistics.  Consequently, scalar WIMPs will result in scalar $s$-wave WIMPonium states.  Majorana fermions are required by their statistics to be in a scalar spin state.  Dirac fermions could potentially
form either particle-particle (antiparticle-antiparticle) bound states (which would be required to be scalar) or particle-antiparticle bound states, which could be scalars or vectors.  A $2p$ state would allow a complex scalar to form a particle-antiparticle bound state which is a vector.
We will often use the usual spectroscopic notation $^{2S+1}L_J$ to label the angular momentum configurations
of the lowest lying bound states.

\section{WIMPonium Effective Theory}
\label{sec:couplings}

We compute the rates for WIMPonium production and decay by using non-relativistic quantum field theory,
borrowing well known results from the study of heavy quarkonium \cite{Bodwin:1994jh}.   We make use of the
fact that for weak coupling, $\alpha_\chi \ll 1$, the velocity of a WIMP inside a WIMPonium state is of order
$v \sim \alpha_\chi$, with good separation between the rest energy $M$, the typical WIMP momentum, 
$M v \sim M \alpha_\chi$, and its kinetic energy, $M v^2 \sim M \alpha^2_{\chi}$.  WIMPonium properties
may thus be thought of as a perturbative expansion in both couplings and velocity.

\subsection{WIMP Interactions with the Standard Model}

Before considering the effective field theory for energies much less than $M$, we consider the relativistic effective
theory at scales of order $M$ which describes the WIMP interactions with SM fields.  These
interactions typically appear as higher dimensional operators involving a pair of WIMPs (to insure the WIMP is stable with respect to decays into SM states) and
a local product of SM fields.  Such contact interactions are typically generated from a UV complete theory by integrating out heavier degrees of freedom.  For example, in the MSSM a 
neutralino-neutralino-$e$-$\bar{e}$ interaction arises from integrating out the selectrons.  As long as the masses
of these additional states are somewhat heavier than that of the WIMPs themselves, this effective field theory
should be a reasonably accurate description of the SM interacting with pairs WIMPs.  It is relatively
straight-forward to generalize our results here to the case in which there are additional dark sector states
close in mass to the WIMP (as will be inevitable, {\em e.g.} in a scenario employing co-annihilation to arrange
for the correct relic density) by explicitly including such states in the effective theory.

A pair of complex scalar WIMPs can interact with Standard Model fields through operators (up to dimension six),
\bea
{\cal L} & = & \lambda |\chi|^2 | H |^2 + \sum_f  \left\{ \frac{y_f}{\Lambda_f^2} |\chi|^2 H \bar{f}_L f_R 
+ \frac{1} {\Lambda_{f_R}^2} \left( \chi^* 
\overleftrightarrow{\partial}_\mu \chi \right) \left[ \bar{f}_R \gamma^\mu f_R \right]
+ \frac{1} {\Lambda_{f_L}^2} \left( \chi^* \overleftrightarrow{\partial}_\mu \chi \right) 
\left[ \bar{f}_L \gamma^\mu f_L \right] \right\}
\nonumber \\ & &
+ \frac{1}{\Lambda_{H4}^2} |\chi|^2 |H|^4
+ \frac{1} {\Lambda_{DH}^2} \left( \chi^* \overleftrightarrow{\partial}_\mu \chi \right) \left( H^\dagger D^\mu H \right)
+\frac{1}{\Lambda^2_W} |\chi|^2 W_{\mu \nu} W^{\mu \nu} 
+ \frac{1}{\Lambda^2_B} |\chi|^2 B_{\mu \nu} B^{\mu \nu}
+ H.c. 
\label{eq:scalarchi}
\eea
where the parameters $\Lambda_i$ and $\lambda$ represent combinations of masses and couplings in the UV theory,
and $H$ is the SM Higgs doublet, including the would-be Goldstone bosons.  Some of these interactions were
previously considered (in a different context) in \cite{Beltran:2008xg}.
We have assumed that there are no new sources of chiral flavor symmetry breaking, and thus the operators
containing $\bar{f}_L f_R$ are flavor conserving and accompanied by the fermion Yukawa interaction $y_f$.
We have assumed the complex scalar WIMP has a conserved global $U(1)_\chi$ symmetry (which could be invoked to motivate the fact that the WIMP stable).  If such a symmetry is
absent, one can also write terms (with independent coefficients)
with $|\chi|^2 \rightarrow \chi^2$.  A real scalar WIMP has only such terms,
and in addition, the terms containing $\chi^* {\partial}_\mu \chi$ may be expressed in terms of
other terms already present by integrating by parts and applying the equations of motion \cite{Dobrescu:2007ec}.

A Dirac WIMP has potentially both scalar and vector interactions with SM fields (up to operators of dimension seven),
\bea
{\cal L} & = & \sum_f \left\{ \frac{1}{\Lambda^2_{f_L}} \left[ \bar{\chi} \gamma^\mu \chi \right]
\left[ \bar{f}_L \gamma_\mu f_L \right]
+ \frac{1}{\Lambda^2_{f_R}} \left[ \bar{\chi} \gamma^\mu \chi \right]
\left[ \bar{f}_R \gamma_\mu f_R \right]
+ \frac{y_f}{\Lambda_f^3} \left[ \bar{\chi} \chi \right] H \bar{f}_L f_R \right\}
+ \frac{1}{\Lambda_H}  \left[ \bar{\chi} \chi \right] | H |^2 
\nonumber \\ & & 
+ \frac{1}{\Lambda^2_{DH}}  \left[ \bar{\chi} \gamma^\mu \chi \right] \left( H^\dagger D_{\mu} H \right)
+ \frac{1}{\Lambda_{H4}^3} \left[ \bar{\chi} \chi \right] |H|^4
+ \frac{1}{\Lambda_{\partial B}^2} \left[ \bar{\chi} \gamma^\mu \partial^\nu \chi \right] B_{\mu \nu}
+ \frac{1}{\Lambda^3_W} \left[ \bar{\chi} \chi \right] W_{\mu \nu} W^{\mu \nu}
\nonumber \\ & & 
+ \frac{1}{\Lambda^3_B} \left[ \bar{\chi} \chi \right] B_{\mu \nu} B^{\mu \nu}
+H.c.
\label{eq:diracchi}
\eea
There are also pseudoscalar ($\bar{\chi} \gamma_5 \chi$) or pseudovector 
($\bar{\chi} \gamma^\mu \gamma_5 \chi$) versions of each operator.  A Majorana WIMP expressed as a
four component spinor looks much the same,
though the $\bar{\chi} \gamma^\mu \chi$ structure vanishes when the WIMP is Majorana.

\subsection{Non-relativistic Effective Theory}

To construct the non-relativistic effective theory, we express the WIMP in terms of a field which annihilates particles
and one which creates antiparticles, with the large time oscillatory behavior due to the mass factored out,
\bea
\chi & \rightarrow & \frac{1}{\sqrt{2 M}} \left[e^{-i M t} \xi + e^{i M t} \eta \right] \\
\chi & \rightarrow & \left[ \begin{array}{c}
e^{-iMt} \xi + i e^{iMt} \frac{\vec{\sigma} \cdot \vec{\nabla}}{2 M} \eta \\
e^{iMt} \eta - i e^{-iMt}  \frac{\vec{\sigma} \cdot \vec{\nabla}}{2 M} \xi
\end{array}
\right]
\eea
where $\xi$ annihilates a WIMP and $\eta$ creates an anti-WIMP, and
the two lines refer to a scalar or fermion WIMP, respectively.  In the fermion WIMP case the
non-relativistic fields are two component Pauli spinors and
$\vec{\sigma}$ are the Pauli matrices.  For a real scalar WIMP
$\eta \rightarrow \xi^*/\sqrt{2}$ (and $\xi \rightarrow \xi/\sqrt{2}$)
and for a Majorana fermion,  $\eta \rightarrow \xi^c / \sqrt{2}  \equiv -i \sigma^2 \xi^* / \sqrt{2}$
(and $\xi \rightarrow \xi / \sqrt{2}$).  These
decompositions lead to kinetic terms in the NR limit in which the fast oscillations have been explicitly removed,
\bea
{\cal L}_{kin} & \rightarrow & \xi^\dagger  \left( i \partial_t + \frac{\nabla^2}{2M} \right) \xi
+  \eta^\dagger \left( i \partial_t - \frac{\nabla^2}{2M} \right) \eta
\eea
with some mild abuse of notation since for a scalar WIMP these are scalar fields, and for a fermion they are Pauli
spinor fields.  Higher time derivatives are explicitly removed by use of the equations of motion, and result in higher order (in $1/M$) relativistic corrections to propagation \cite{Caswell:1985ui}.

WIMPonium production and decay are both captured by the imaginary coefficients of local four-WIMP operators which arise in the UV theory as ``box" diagrams in which two WIMPs annihilate into SM fields which then recombine into two WIMPs.  In terms of the couplings defined in equations~(\ref{eq:scalarchi}) and (\ref{eq:diracchi}) they generally arise as two applications of a given term with the Standard Model fields contracted into a loop.  

For a scalar WIMP, the relevant operators for WIMPonium decay take the form,
\bea
 \sum_{i} \frac{C^{i}_{f\bar{f}}(^1S_0)+ C^{i}_{HH}(^1S_0) + C^{i}_{WW}(^1S_0) + C^{i}_{BB}(^1S_0)}{M^2}{\cal O}^i(^1S_0)
+ \frac{C_{f\bar{f}}(^1P_1) + C_{HH}(^1P_1)}{M^4} {\cal O}(^1P_1),
\label{eq:sdecay}
\eea
where we have included the intermediate states $f \bar{f}$, $H^*H$, $WW$, and $BB$. The four-WIMP
operators encoding WIMPonium decays are,
\bea
{\cal O}^+(^1S_0) & = & \left( \xi^*\xi^*\right)| 0 \rangle \langle 0 | \left( \xi\xi \right) ,\\
{\cal O}^-(^1S_0) & = & \left( \eta \eta \right)| 0 \rangle \langle 0 | \left( \eta^* \eta^* \right) ,\\
{\cal O}^0(^1S_0) & = & \left( \eta \xi^* \right) | 0 \rangle \langle 0 | \left( \eta^* \xi \right) ,\\
{\cal O}(^1P_1) & = & \left( \eta^* \vec{\nabla} \xi \right) | 0 \rangle \langle 0 | \left( \xi^* \vec{\nabla} \eta \right),
\eea
corresponding to $s$-wave particle-particle, antiparticle-antiparticle, and particle-antiparticle states and a $p$ wave particle-antiparticle state, respectively.  The Wilson coefficients $C_i$ may be computed in terms of the interactions of
Eq.~(\ref{eq:scalarchi}) and will have imaginary parts (through the optical theorem) when a pair of WIMPs is able to non-relativistically annihilate into the given channel.  There will also often be annihilations into the light particle responsible for binding the WIMPonium system which may make up an appreciable branching fraction.

WIMPonium production is described by similar operators.  For a scalar WIMP, we restrict ourselves to consideration of unpolarized production of the $^1P_1$ state $\vec{\Psi}_m$ (where $m=-1,0,1$ is the spin index)
through $f \bar{f}$ fusion at leading order in the velocity expansion,
\bea
\frac{D_{f \bar{f}} (^1P_1 )}{M^5} 
\left( \eta^\dagger \vec{\nabla} \xi \right) \left( \sum_{X,m} | \vec{\Psi}_m + X \rangle \langle \vec{\Psi}_m + X| \right)
\left( \xi^\dagger \vec{\nabla} \eta \right),
\eea
where the sum over $X$ is over all SM states in addition to the WIMPonium itself, and the production coefficient 
$D_{f \bar{f}} (^1P_1 )$ (which has dimensions of $1/[{\rm Mass}]^2$) depends on the event kinematics associated with the WIMPonium and/or other SM particles involved in the reaction.

Similar, a fermion WIMP has operators representing its decays,
\bea
 \sum_{i} \frac{C^{i}_{f\bar{f}}(^1S_0)+ C^{i}_{HH}(^1S_0) + C^{i}_{WW}(^1S_0) 
 + C^{i}_{BB}(^1S_0)}{M^2}{\cal O}^i(^1S_0)
+ \frac{C_{f\bar{f}}(^3S_1) + C_{HH}(^3S_1)}{M^2} {\cal O}(^3S_1),
\label{eq:fdecay}
\eea
with
\bea
{\cal O}^+(^1S_0) & = & \left( \xi^\dagger \xi^c\right)| 0 \rangle \langle 0 | \left( \xi^{c\dagger} \xi \right), \\
{\cal O}^-(^1S_0) & = & \left( \eta^{c\dagger} \eta \right)| 0\rangle \langle 0 | 
\left( \eta^\dagger \eta^c \right) ,\\
{\cal O}^0(^1S_0) & = & \left( \xi^\dagger \eta \right) | 0 \rangle \langle 0 | \left( \eta^\dagger \xi \right) ,\\
{\cal O}(^3S_1) & = & \left( \eta^\dagger \vec{\sigma} \xi \right) | 0 \rangle \langle 0 | 
\left( \xi^\dagger \vec{\sigma} \eta \right),
\eea
and $p$-wave state contributions are sub-leading in the velocity expansion.  Again, restricting ourselves to unpolarized production
of a vector WIMPonium state from an $f \bar{f}$ initial state, the relevant operator for production is,
\bea
\frac{D_{f \bar{f}} (^3S_1 )}{M^3}
\left( \eta^\dagger \vec{\sigma} \xi \right) \left( \sum_{X,m} | \Psi_m + X \rangle \langle \Psi_m + X| \right)
\left( \xi^\dagger \vec{\sigma} \eta \right)~.
\eea

\subsection{Matrix Elements}

In applying our effective theory to WIMPonium production and decay processes, we encounter
matrix elements of the non-relativistic WIMP fields sandwiched between WIMP vacua $|0\rangle$ and 
WIMPonium states $| \Psi \rangle$.  These matrix elements can be expressed at leading order in the velocity expansion in terms of a regularized WIMPonium radial wave function $\psi_r$ and its derivative $\psi^\prime_r$ at the origin \cite{Berger:1980ni,Bodwin:1994jh},
\bea
\langle \Psi | {\cal O}(^1S_0) | \Psi \rangle & \simeq & 
\langle \Psi | {\cal O}(^3S_1) | \Psi \rangle \simeq \frac{1}{2\pi} \overline{| \psi_r|^2} 
\sim \left( \alpha_\chi M \right)^3 \\
\langle \Psi | {\cal O}(^1P_1) | \Psi \rangle & \simeq & \frac{3}{2\pi} \overline{| \psi^\prime_r |^2}
\sim \left( \alpha_\chi M \right)^5
\eea
where the estimates for the sizes of the matrix elements follow from the velocity expansion
\cite{Lepage:1992tx}, and are appropriate for WIMPonium momentum eigenstates normalized to volume.

\section{Implications at Future Colliders}
\label{sec:collider}

\subsection{Vector WIMPonium}

Vector WIMPonium states may result in theories of complex scalar (as $^1P_1$ states) or Dirac fermion
(as $^3S_1$ states) dark matter.  At colliders, unlike open WIMP production, which leads to signals of missing energy, vector WIMPonium signals are more like $Z^\prime$ searches, leading to resonance signals
of SM particles.
WIMPonium decay widths into a final state $Y$ are obtained at leading order in the velocity expansion from the imaginary parts of the Wilson coefficients in the four WIMP operators of Equation~(\ref{eq:sdecay}) or (\ref{eq:fdecay}),
\bea
\Gamma ( \vec{\Psi} \rightarrow Y ) & = & \frac{2~{\rm Im}~ C_Y(^1P_1)}{M^4} 
\left| \langle 0 | \eta^* \vec{\nabla} \xi | \vec{\Psi} \rangle \right|^2
\sim 2\alpha_\chi^5M~{\rm Im}~C_Y (^1P_1)
\eea
for a scalar WIMP in a $p$-wave state, and
\bea
\Gamma ( \vec{\Psi} \rightarrow Y ) & = & \frac{2~{\rm Im}~ C_Y(^3S_1)}{M^2} 
\left| \langle 0 | \eta^\dagger \vec{\sigma} \xi | \vec{\Psi} \rangle \right|^2
\sim 2\alpha_\chi^3M~{\rm Im}~C_Y (^3S_1)
\eea
for a Dirac fermion WIMP in an $s$-wave state.  For the case of decays into light SM fermions
($m_f \ll M$), the Wilson Coefficient is given by,
\bea
{\rm Im}~C_{f \bar{f}}(^1P_1) =  {\rm Im}~C_{f \bar{f}}(^3S_1) & = & N_f~\frac{1}{24 \pi}~\left(
\frac{M^4}{\Lambda^4_{f_R}} + \frac{M^4}{\Lambda^4_{f_L}}
\right)
\eea
where $N_f$ is the number of colors of $f$, three for quarks, and one for leptons.  These widths are generally very small, but nonetheless WIMPonium is still expected to decay promptly on collider time scales.

Production of an unpolarized $\vec{\Psi}$ at hadron colliders can be driven by $q \bar{q} \rightarrow \vec{\Psi}$, which at leading order involves the same matrix elements as the WIMPonium decays,
\bea
\sigma (q \bar{q} \rightarrow \vec{\Psi} ) & = & \frac{D_{f\bar{f}}(^1P_1)}{M^5} \sum_{X,m}
\langle0 | \left( \eta^\dagger \vec{\nabla} \xi \right)  | \vec{\Psi}_m + X \rangle \langle \vec{\Psi}_m + X|
\left( \xi^\dagger \vec{\nabla} \eta \right) | 0 \rangle 
\sim \alpha^5_\chi ~D_{f\bar{f}}(^1P_1)
\label{eq:vecprods}
\eea
for a scalar WIMP and
\bea
\sigma (q \bar{q} \rightarrow \vec{\Psi} ) & = & \frac{D_{f\bar{f}}(^3S_1)}{M^3} \sum_{X,m}
\langle0 | \left( \eta^\dagger \vec{\sigma} \xi \right)  | \vec{\Psi}_m + X \rangle \langle \vec{\Psi}_m + X|
\left( \xi^\dagger \vec{\sigma} \eta \right) | 0 \rangle 
\sim \alpha^3_\chi ~D_{f\bar{f}}(^3S_1)
\label{eq:vecprodf}
\eea
for a fermion WIMP, where at leading order in QCD,
\bea
D_{f \bar{f}} & = & \frac{\pi}{48 s} \left( \frac{M^4}{\Lambda^4_{f_R}} + \frac{M^4}{\Lambda^4_{f_L}}  \right)
~ \delta \left( 1 - \frac{M_{\Psi}^2}{s} \right)
\eea
where $s$ is the Mandelstam variable corresponding to the partonic center of mass energy squared
and we have used the fact that to good approximation, $M_\Psi \sim 2 M$.
These ``partonic" cross sections are convolved with the parton distribution functions
as usual in order to obtain the hadro-production cross sections.

With a specific model (or specific choices of the $\Lambda_f$s), we can make predictions for production rates
and decay BRs of vector WIMPonium.  For simplicity, we set $\Lambda_{f_R} = \Lambda_{f_L} = \Lambda$
equal for all fermions, and switch off the interactions with all other Standard Model fields.  We can obtain
Tevatron limits from $Z^\prime$ searches \cite{Abulencia:2005nf} in the language of the $c_u$ and $c_d$
parameters \cite{Carena:2004xs} (which are equal under the assumptions that fermions all couple universally).
For universal fermion couplings and no other decay modes, we have 
\bea
c_u, c_d  \sim \frac{4}{45} \alpha_\chi^5 ~~~ ({\rm scalar~WIMP})& ~~~~~~~~~~~~~ & 
c_u, c_d  \sim \frac{4}{45} \alpha_\chi^3 ~~~({\rm fermion~WIMP})
\eea
where the prefactors assume decays into top quarks are allowed and
these can be interpreted as upper limits since there are probably fairly large branching fractions into
the mediator states whose signatures are model-dependent.  The CDF limits on $c_u$ and $c_d$
range from a few times $10^{-5}$ at resonance masses of 150~GeV to about $10^{-2}$ at TeV masses.  In terms of
$\alpha_\chi$, this requires $\alpha_\chi \lesssim 0.2~(0.06)$ for a scalar (fermion) WIMP whose mass is 
$M = 75$~GeV and allows essentially any perturbative $\alpha_\chi$ for WIMPS of mass $M=500$~GeV.

The operators $\left( \chi^* \overleftrightarrow{\partial}_\mu \chi \right) \left( H^\dagger D^\mu H \right)$
and $\left[ \bar{\chi} \gamma^\mu \chi \right] \left( H^\dagger D_{\mu} H \right)$ will lead, after electroweak
symmetry-breaking, to mass mixing between a vector WIMPonium state and the $Z$ boson, with the
effective mass$^2$ matrix taking the form in the $Z$-$\vec{\Psi}$ basis,
\bea
\frac{1}{2}
\left(
\begin{array}{cc}
M_Z^2 & \alpha^{n}_\chi \frac{g_Z v^2}{4 \Lambda^2_{DH}} M^2_{\Psi} \\
 \alpha^{n}_\chi \frac{g_Z v^2}{4 \Lambda^2_{DH}} M^2_{\Psi}  &  M_{\Psi}^2
\end{array}
\right)
\eea
where $n=5/2$ for a scalar WIMP and $n=3/2$ for a fermion WIMP.
Diagonalizing this matrix (in the limit $M_Z \ll M_\Psi$) leads to $Z$-$\vec{\Psi}$ mixing of the order of
$\alpha_{\chi}^{2n} M_Z^2 / (4 g_Z^2 \Lambda^2_{DH})$.
Electroweak precision data generically constrains the properties of the $Z$ boson to lie within one part per
mil or so of the SM values \cite{:2003ih}.  
The operator $\left[ \bar{\chi} \gamma^\mu \partial^\nu \chi \right] B_{\mu \nu}$ leads to kinetic mixing between
$\vec{\Psi}$ and the hyper-charge boson, which is much less constrained than mass mixing.  In the absence
of any other WIMP couplings to the Standard Model from the UV theory, such mixing may provide the dominant contributions to
vector WIMPonium production and decay.

At the LHC, $\vec{\Psi}$ appears like a weakly coupled $Z^\prime$ provided it has appreciable decay into
$\ell^+ \ell^-$ pairs.  The cross section will be suppressed by $\alpha_\chi^3$ (fermion WIMP) or
$\alpha_\chi^5$ (scalar WIMP), and thus is expected to be much smaller than open WIMP production.  However,
depending on the underlying theory, the resonance in lepton pairs may in fact turn out to be easier to pick
out from the background than the missing energy signatures of open WIMP production.

\begin{figure}
\hspace*{2.5cm}
\includegraphics[angle=270,scale=0.5]{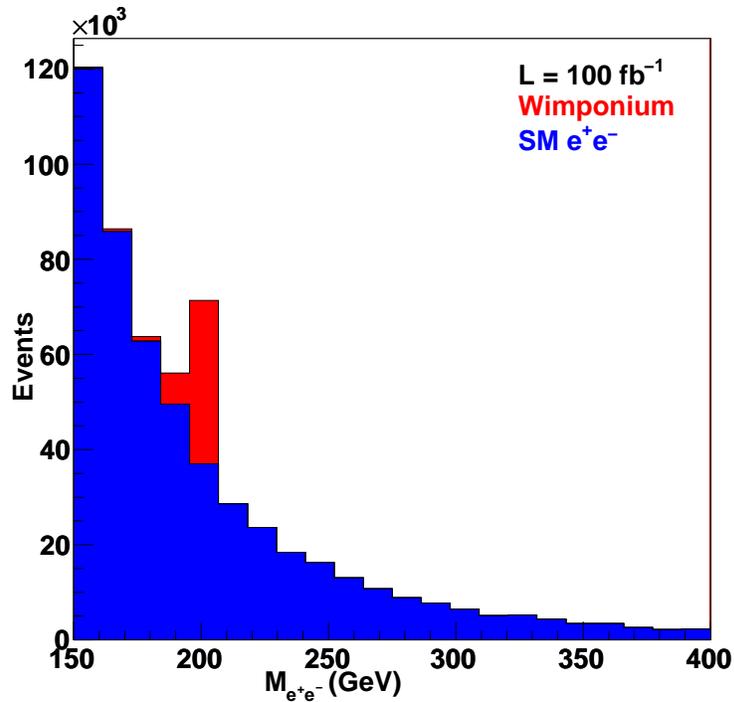} 
\vspace*{-1cm}
\caption{\label{fig:mll}
Sample reconstructed invariant mass of the $\ell^+ \ell^-$ for the Standard Model (blue) and a vector WIMPonium
$^3S_1$ state
signal for a mass of 200 GeV, universal $\Lambda_f = M$, and $\alpha_\chi = 0.2$ (red) at the LHC for
an integrated luminosity of 100 fb$^{-1}$.  Note that the intrinsic width of the WIMPonium and the experimental
resolution are much smaller than the bin width.
}
\end{figure}

In order to simulate the WIMPonium signal, we reproduce the cross sections at the LHC in equations~(\ref{eq:vecprods}) and (\ref{eq:vecprodf}) using MadEvent \cite{Alwall:2007st} to produce 
$pp \rightarrow \vec{\Psi} \rightarrow \ell^+ \ell^-$ and SM background, which we reconstruct with the default LHC detector simulation of PGS \cite{PGS}.
An example reconstructed invariant mass distribution of $\ell^+ \ell^-$ for signal and background for a $M=100$~GeV Dirac fermion WIMP with universal WIMP couplings to
fermions $\Lambda_f = M$ and $\alpha_\chi \sim 0.1$ is shown in Figure~\ref{fig:mll}.

\begin{figure}
\hspace*{.5cm}
\includegraphics[angle=0,scale=0.5]{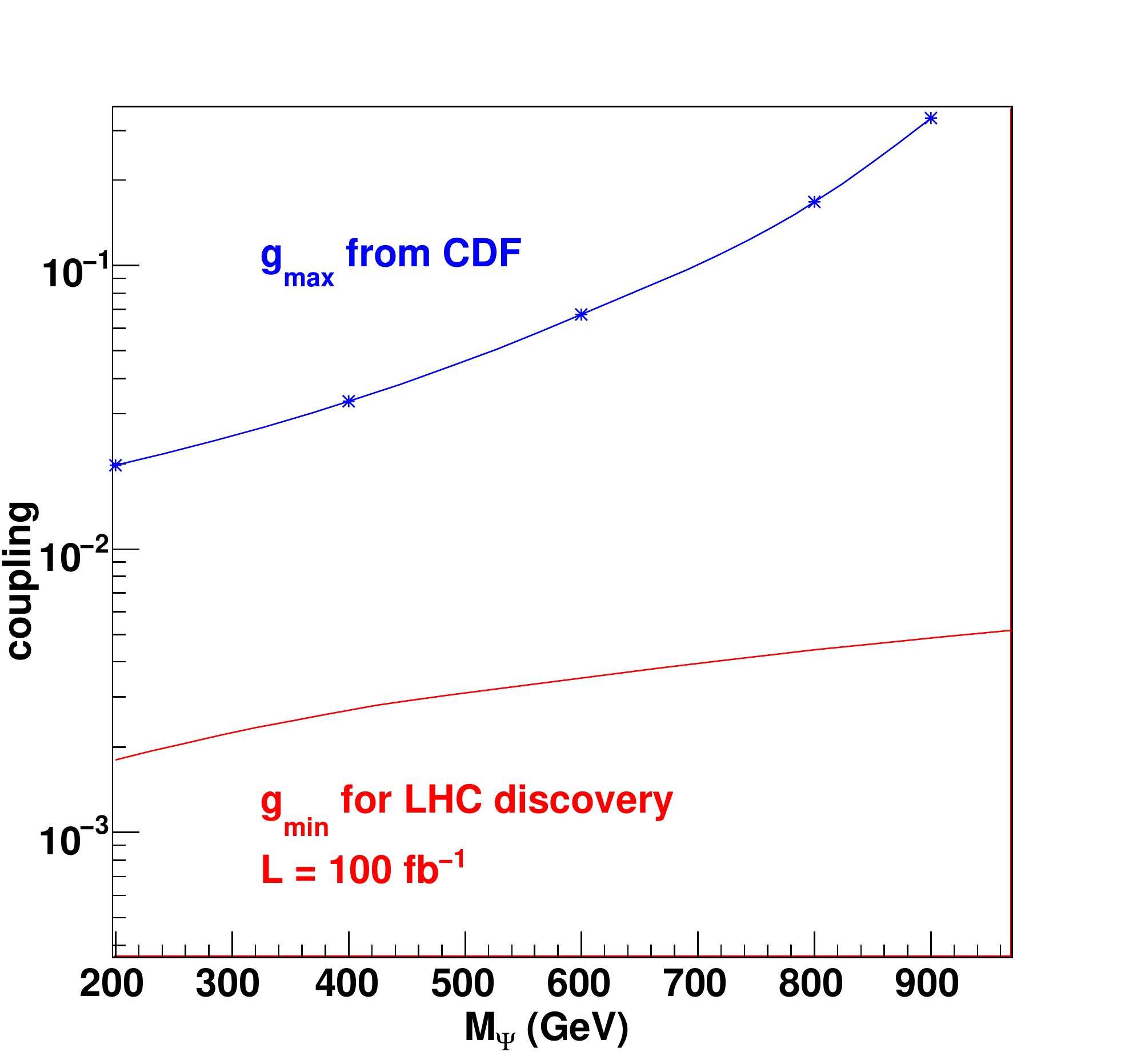} 
\caption{\label{fig:gmin}
Minimum coupling for a weakly coupled vector WIMPononium state to be visible at $5\sigma$ against the SM
background in the $\ell^+ \ell^-$ channel as a function of the WIMPonium mass, assuming universal coupling
to all SM fermions, no coupling to SM bosons, and no non-SM decay modes (red curve).  Also shown are
the CDF limits from $Z^\prime$ searches as a function of mass (blue points and interpolating curve).
}
\end{figure}

We determine the minimum universal effective coupling of vector WIMPonium to fermions (continuing to assume
universal $\Lambda_f = M$ and no other open decay modes) for a $5\sigma$ discovery of the WIMPonium
against the SM background (in an invariant mass cut window of $\pm 10$~GeV -- well above both the intrinsic WIMPonium width and the expected experimental leptonic invariant mass resolution -- around the assumed mass of the WIMPonium) at the LHC with 100 fb$^{-1}$.  We plot our results as a function of the WIMPonium mass in
Figure~\ref{fig:gmin}, where the effective coupling is related to the universal $\Lambda$ by,
\bea
g_{eff} \simeq \alpha_\chi^{5/2} \frac{M^2}{\Lambda^2} ~~~ ({\rm scalar~WIMP})& ~~~~~~~~~~~~~ & 
g_{eff} \simeq  \alpha_\chi^{3/2} \frac{M^2}{\Lambda^2} ~~~({\rm fermion~WIMP})
\eea

Once a WIMPonium state is discovered, it can reveal a lot about the underlying theory
of dark matter.  For one thing, the mass of a WIMPonium state decaying into leptons can be measured
with great precision, in contrast to mass measurements of WIMPs in missing energy events, which are
expected to lead to far larger uncertainties.  Further detailed study of WIMPonium branching ratios and
production/ decay distributions can reveal many of the details of the pattern of couplings to the 
SM particles \cite{Petriello:2008zr}.  The detailed couplings can be mapped back into the WIMP-SM couplings of
equations~(\ref{eq:scalarchi}) or (\ref{eq:diracchi}), and from there into the UV theory.

\subsection{Scalar WIMPonium}

Scalar WIMPonium states are very generic.  They can occur for any type of scalar or spin 1/2 WIMP.  In the case
of a complex scalar or Dirac fermion WIMP with a scalar mediator particle, there may be several scalar
states, composed of particle-particle ($\Psi^+$), particle-antiparticle ($\Psi^0$), and antiparticle-antiparticle
($\Psi^-$) systems.  If there is an exact $U(1)_\chi$ symmetry, $\Psi^{\pm}$ will be stable, which can lead to
interesting phenomena in cosmology and astrophysics.  If $U(1)_\chi$ is broken, the three states will mix with each other, and all three will be unstable.

Under our assumption that the dark sector does not contribute any new sources of chiral symmetry-breaking to the SM fermions, scalar WIMPonium is difficult to produce in the laboratory, because it couples very weakly to light fermions.  The leading contributions to decay widths are identical for fermion and scalar WIMPs,
\bea
\Gamma ( \Psi^0 \rightarrow Y ) & = & \frac{2~{\rm Im}~ C_Y(^1S_0)}{M^2} 
\left| \langle 0 | \eta^\dagger \xi | \Psi^0 \rangle \right|^2
\sim 2\alpha_\chi^3M~{\rm Im}~C_Y (^1S_0)
\eea
(where for illustration, we write the result for a Dirac fermion $\Psi^0$ state)
because the leading matrix elements for the $^1S_0$ states are identical in all cases.

Just as a vector WIMPonium state can mix with the $Z$ boson, scalar WIMPonium can mix with the Standard
Model Higgs, and have an impact on its couplings.  Unless $\alpha_\chi$ is very large, it is difficult to
imagine that such mixing would produce an effect large enough to be observable at the LHC
\cite{Zeppenfeld:2002ng}, though effects could easily be visible if the Higgs is accessible to a TeV scale
$e^+ e^-$ linear collider \cite{Battaglia:2000jb}.

\section{Implications for Cosmology and Particle Astrophysics}
\label{sec:astro}

As alluded to in the introduction, the excesses recently reported by the PAMELA and ATIC experiments have
inspired exploration of Sommerfeld enhancements in order to motivate the large annihilation cross sections required by dark matter interpretations of these signals \cite{ArkaniHamed:2008qn,Pospelov:2008jd}.
Such large cross sections are seeming at odds with the picture of the WIMP as a thermal relic, and require the
effective cross section to be smaller at the time of freeze-out than today.  This is accomplished by Sommerfeld corrections, which are subdominant at the time of freeze-out, when WIMP velocities are on the order of
$v \sim 1/5$, but grow at later times when WIMP kinetic energies become comparable to the long range potential.
This enhancement cuts off for very low velocities, once exchanged momenta are of order the mass $m$ of the mediator.  There is the potential for large effects on the expected rate from indirect DM searches in present day haloes \cite{ArkaniHamed:2008qn,Pospelov:2008jd,MarchRussell:2008tu,Kamionkowski:2008gj}.

In addition to boosting the rate of traditional indirect DM searches, the presence of WIMPonium states can also lead to new astrophysical phenomena as well.  WIMPs scattering in haloes may form WIMPonium by emitting a boson (on or off-shell) with energy on order the binding energy of the WIMPonium.  As we saw in 
section~\ref{sec:wimponium}, the binding energy for TeV mass WIMPs will typically be of order MeV (up to GeV), and can produce $e^+ e^-$ pairs (perhaps contributing the the INTEGRAL/SPI excess \cite{Weidenspointner:2006nua})
as well as gamma rays.  If higher radial modes are available, there could be a whole forest of such lines produced by radiative transitions between the various states. Transitions will particularly be favored to direct decay through WIMP annihilation if the WIMPonium is produced in a higher angular momentum state \cite{MarchRussell:2008tu}.
While such lines are probably challenging to detect, their detailed spectroscopy encodes the physics of the Yukawa potential responsible for binding the WIMPs, which as we saw above determines the number and energy spacings of the bound states.

If the $\Psi^{\pm}$ states are stable, some fraction of WIMPs in haloes will probably become bound into them.  Since these states don't decay through annihilation, they will persist for long times, and can enhance transitions between different energy levels.  The chemical equilibrium between these stable states and free WIMPs could also play a role in the freeze-out process of the WIMP and resultant relic density.  Such states could also make up some fraction of the WIMPs expected to collect in the Earth and Sun.  We leave detailed exploration of such phenomena for future work.

\section{Outlook}
\label{sec:outlook}

While very simple models of dark matter are appealing, the true dark sector may turn out to be non-minimal
and involve interesting inter-WIMP dynamics, which may prove important in interpreting signals at future colliders or from astroparticle searches for dark matter.  In this article we have explored some of the interesting consequences of one such non-minimal dark sector -- the possibility that WIMPs may have weakly bound states, or WIMPonium.  Such bound states are a smoking-gun signature of the dynamics responsible for a Sommerfeld enhancement of WIMP annihilation, which has recently received much theoretical attention because of the experimental results from PAMELA and ATIC.  We have further demonstrated that WIMPonium is very interesting in its own right.

We have constructed effective field theories to describe WIMP interactions with the Standard Model, allowing us to be agnostic as the underlying nature of the model of dark matter.  We went on to
use the language of NRQCD (as usually applied to heavy quarkonia systems) to explore the bounds from LEP and the Tevatron, and the prospects for future detection at the LHC.  While we have restricted ourselves to leading order predictions in the WIMP velocity expansion and order of magnitude estimates for the matrix elements of operators, the formalism allows these approximations to be systematically improved.

The properties of WIMPonium reflect many of the underlying properties of the WIMPs themselves, and WIMPonium states can leave their trace at the LHC or by influencing the relic density or signals of indirect detection of dark matter.  The identity of the dark matter is still a mystery.  There is no doubt that unravelling that mystery will teach us a lot, and reveal the elegant dynamics of dark matter.

\vspace*{1cm}
{\bf Acknowledgments }\\

The authors are pleased to acknowledge conversations with E.~Berger, D.~Hooper, I.~Low, A.~Pierce, 
M.~Schmitt, T.~Rizzo,
X. Tormo and especially G.~Bodwin.
Research at Argonne National Laboratory is 
supported in part by the Department of Energy 
under contract DE-AC02-06CH11357.  T~Tait is grateful to the 
SLAC theory group for his many visits, and to the KITP for providing an excellent environment
in which some of this work was accomplished.  That portion of the research was
supported in part by the National Science Foundation under
Grant No. PHY05-51164.\\


\end{document}